\documentclass[twocolumn]{webofc}

\usepackage[varg]{txfonts}   
\usepackage{amsmath}

\begin{document}
\title{Three-body description of $\boldsymbol{^{12}}$C: From the hyperspherical formulation to the algebraic cluster model and its application to $\boldsymbol{\alpha}+\boldsymbol{^{12}}$C inelastic scattering}
%
%

\author{\firstname{J.} \lastname{Casal}\inst{1}\fnsep\thanks{\email{casal@pd.infn.it}} \and
        \firstname{L.} \lastname{Fortunato}\inst{1}\and
        \firstname{A.} \lastname{Vitturi}\inst{1}\and
        \firstname{E. G.} \lastname{Lanza}\inst{2}
}

\institute{Dipartimento di Fisica e Astronomia ``G. Galilei'' and INFN Sezione di Padova, I-35131 Padova, Italy.
\and
           INFN Sezione di Catania, I-95123, Catania, Italy
          }

\abstract{%
  Form factors for $\alpha+{^{12}}$C inelastic scattering are obtained within two theoretical ($\alpha+\alpha+\alpha$) approaches: The hyperspherical framework for three identical bosons, and the algebraic cluster model assuming the $D_{3h}$ symmetry of an equilateral triangle subject to rotations and vibrations. Results show a good agreement, with form factors involving the Hoyle state having a slightly larger extension within the hyperspherical approach. Coupled-channel calculations using these form factors are ongoing.
}
\maketitle
\section{Introduction}
\label{intro}
The low-lying structure of $^{12}$C is still one of the most fascinating open problems in nuclear physics. Alpha-clusterization and the nature of the so-called Hoyle state, which plays a crucial role in nucleosynthesis, have attracted special interest. Microscopic theories (e.g., \cite{Chernykh07}) support the existence of three-alpha cluster configurations for the $^{12}$C nucleus, a fact which justifies the use of cluster models (e.g.,~\cite{Nguyen13}) and algebraic methods (e.g.,~\cite{dellaRocca17}). These approaches, although simpler, are particularly suitable for the description of reaction observables. 
Experimentally, different probes have been extensively used to access the properties of its ground and excited states. We discuss here the case of $\alpha+{^{12}}$C inelastic scattering studied within two theoretical three-body approaches: The hyperspherical formalism, and the algebraic cluster model. 
Our goal is to compare form factors for inelastic scattering in these two approaches and set the basis for full coupled-channel calculations.
\section{Three-body calculations}
\label{sec-1}
As in Refs.~\cite{Nguyen13,Ishikawa14}, the $\alpha+\alpha+\alpha$ problem can be solved within the hyperspherical formalism~\cite{Zhukov93}, which has been successfully applied to describe the properties of Borromean nuclei such as the two-neutron halo systems $^{6}$He or $^{11}$Li, or the weakly bound stable nucleus $^{9}$Be~\cite{JCasal14}. Within this framework, the states of the three-body system can be written as
\begin{equation}
\Psi^{j\mu}(\rho,\Omega)=\rho^{-5/2}\sum_{\beta}\chi_\beta^j(\rho)\Upsilon_\beta^{j\mu}(\Omega),
\label{eq:wf}
\end{equation}
where $\rho$ is the hyper-radius, $\Omega$ contains all the angular dependence, functions $\Upsilon_\beta^{j\mu}$ are the so-called hyperspherical harmonics, and $\chi_\beta^j$ are the hyperradial functions to be determined. Here, $\beta$ is the set of quantum numbers defining each channel. In the case of three zero-spin particles, $\beta\equiv\{K,l_x,l_y\}$, where $l_x$ and $l_y$ are the orbital angular momenta associated to the usual Jacobi coordinates, 
so that $\boldsymbol{j}=\boldsymbol{l_x}+\boldsymbol{l_y}$ gives the total angular momentum, and $K$ is the so-called hypermomentum. Note that $l_x$ has to be even for identical bosons. Details can be found in Ref.~\cite{Nguyen13}. The hyperangular functions and $\chi_\beta^j(\rho)$ can be expanded in a given basis,
\begin{equation}
\chi_{\beta}^{j}(\rho) = \sum_{i} C_{i\beta}^{j} U_{i\beta}(\rho).
\end{equation}
Index $i$ runs over the number of hyperradial excitations included, and coefficients $C_{i\beta}^j$ are obtained upon diagonalization of the three-body Hamiltonian with coupling potentials 
\begin{equation}
V_{\beta'\beta}^{j\mu}(\rho)=\left\langle \Upsilon_{\beta }^{j\mu}(\Omega)\Bigg|\sum_{j>i=1}^3 V_{ij}(\rho,\alpha) \Bigg|\Upsilon_{\beta'}^{ j\mu}(\Omega) \right\rangle + \delta_{\beta,\beta'}V_{3b}^j(\rho).
\end{equation}
This contains all binary interactions, integrated over the angular dependence, and a phenomenological three-body force to account for effects not explicitly included in the binary interactions~\cite{IJThompson04}.
In this work, calculations are performed with the Ali-Bodmer $\alpha$-$\alpha$ nuclear potential~\cite{AliBodmer} together with a hard-sphere Coulomb term. Radial functions are expanded in a THO basis~\cite{JCasal14}, and the three-body force is adjusted to reproduce the energies of the first $2^+_1$ bound excited state and the $0^+_2$ Hoyle state in $^{12}$C, which have a well-developed three-body character~\cite{Chernykh07}.  



The set of coupled differential equations to describe $\alpha+^{12}\text{C}$ inelastic scattering requires the computation of radial form factors involving the projectile-target interaction. Within the hyperspherical three-body framework, this is a four-body problem involving the $\alpha$-$\alpha$ potential. Between states labeled $i$ and $j$, the form factor is simply
\begin{equation}
F_{ij} (\vec{R}) = \langle \Psi_i|\sum_{q=1}^3 V_{\alpha\alpha}(|\vec{r}_q-\vec{R}|)|\Psi_j\rangle,
\end{equation}
where $\vec{R}$ is the projectile-target distance and $\vec{r}_q$ is the position of each $\alpha$ within the target. For this purpose, the potential we use is derived from the double folding of two Gaussian densities, adjusted to reproduce the radius of the $\alpha$ particle, with the M3Y nucleon-nucleon interaction~\cite{m3y}. This way, the computed form factors will be consistent with those described in the next section within the formalism of densities and transition densities.


\section{Algebraic cluster model}
\label{sec-2}

Assuming the $D_{3h}$ symmetry corresponding to an equilateral triangle, the density for the ground-state band for $^{12}$C can be writte as~\cite{dellaRocca17}
\begin{equation}
\rho_{\rm g.s.} (\vec{r},\{\vec{r}_k\}) = \sum_{k=1}^3 \rho_\alpha (\vec{r}-\vec{r}_k), ~~~~~~ \rho_\alpha (\vec{r}) = \left(\frac{d}{\pi}\right)^{3/2} e^{-dr^2},
\end{equation}
with $d=0.56(2)$ fm$^{-2}$ to reproduce the radius of the $\alpha$ particle and $\{\vec{r}_k\}$ at the vertices of the triangle, i.e., $\vec{r}_1=(\beta,\pi/2,0)$, $\vec{r}_2=(\beta,\pi/2,2\pi/3)$ and $\vec{r}_3=(\beta,\pi/2,4\pi/3)$ in spherical coordinates. The radial parameter $\beta=1.82$ fm ensures the $0^+_1$ ground-state radius and $B(E2)$ value to the first $2^+_1$ state are reproduced. Expanded in spherical harmonics, it takes the form
\begin{equation}
\rho_{\rm g.s.} (\vec{r})=\sum_{\lambda\mu} \rho_{\rm g.s.}^{\lambda\mu}(r) Y_{\lambda\mu}(\theta,\varphi),
\end{equation}
where only the multipoles allowed by the $D_{3h}$ symmetry appear in the sum. The 00 term represents the 0$^+_1$ ground-state density, while others represent the change in density for transitions to higher lying states of the same band, e.g., 20 is the term associated to the $2^+_1$ state.

By considering now symmetric vibrations $\Delta\beta^A$ along the radial direction, we can construct the band associated to the Hoyle state as a breathing mode. The transition densities connecting the ground-state band with this $A$-type band can be obtained in leading order as
\begin{equation}
\delta\rho_{\text{g.s.}\rightarrow A}(\vec{r}) = \chi_1 \frac{d}{d\beta} \rho_{\rm g.s.} (\vec{r},\beta),
\end{equation}
with $\chi_1\simeq 0.247$ to recover the experimental monopole matrix element $M(E0)$. Again, expanding in multipoles one gets
\begin{equation}
\delta\rho_{\text{g.s.}\rightarrow A}(\vec{r}) = \sum_{\lambda\mu} \delta\rho_{\text{g.s.}\rightarrow A}^{\lambda\mu}(r) Y_{\lambda\mu}(\theta,\varphi).
\end{equation}
For details about these densities and transition densities, see Refs.~\cite{Vitturi19,vittprep19}. In this case, form factors for $\alpha+^{12}\text{C}$ inelastic scattering can be obtained following a double folding procedure with the M3Y interaction introduced above, 
\begin{equation}
F_{ij} (\vec{R}) = \int\int \rho_\alpha(\vec{r}_1-\vec{R}) v_{NN}(|\vec{r}_{12}|) \delta\rho^{i\rightarrow j}(\vec{r}_2)d\vec{r}_1d\vec{r}_2.
\end{equation}

\section{Comparison of form factors}
\label{sec-3}

\begin{figure}[h]
	\centering
	\includegraphics[width=\linewidth]{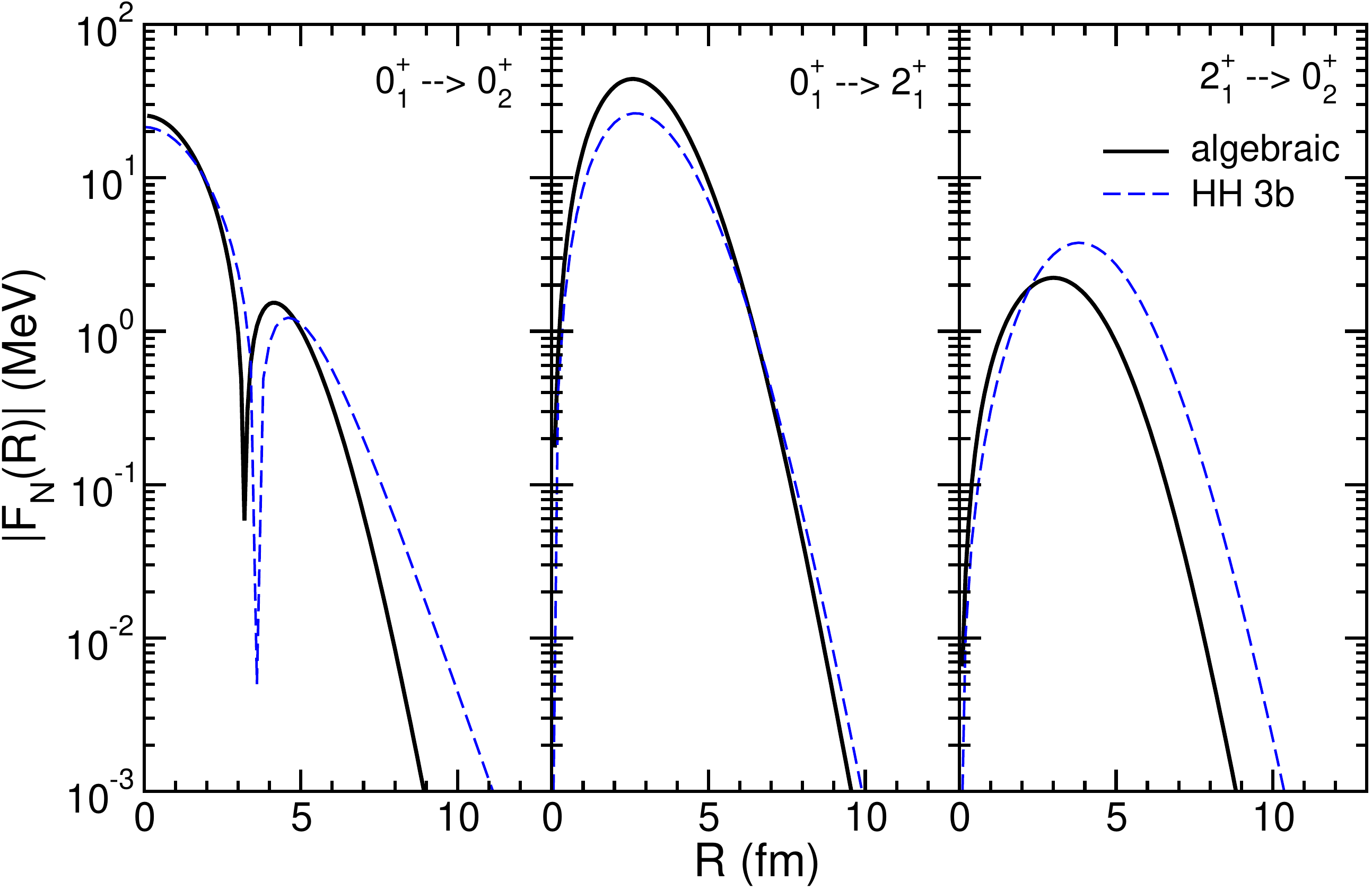}
	\caption{Nuclear form factors for $\alpha+{^{12}}$C inelastic scattering. Comparison between the three-body hyperspherical formalism (dashed) and the algebraic cluster model (solid). See text.}
	\label{fig:ff}
\end{figure}

The form factors for the $0^+_1\rightarrow0^+_2$, $0^+_1\rightarrow2^+_1$ and $2^+_1\rightarrow0^+_2$ transitions, computed within the two approaches, are shown for comparison in Fig.~\ref{fig:ff}. A rather good agreement is observed, even though they come from very different theoretical approaches. This may indicate that the two models capture essentially the same geometrical properties of $^{12}$C. Form factors connecting the bound states ($0^+_1,2^+_1$) with the Hoyle state ($0^+_2$) seem to exhibit a larger extension within the hyperspherical approach. This may have implication for the corresponding cross section. Full coupled-channel calculations involving these form factors, as well as those connecting other low-lying states in $^{12}$C, are ongoing and will be presented elsewhere.

\bibliography{bib12C.bib}

\end{document}